\documentclass[11pt,twoside]{article}
\usepackage{asp2004}
\usepackage{epsf}
\usepackage{psfig}
\usepackage{lscape}
\begin{document}

\title{Polarimetry and the Envelopes of Magellanic B[e] Supergiants}
\author{A.M.Magalh\~aes, R. Melgarejo, A. Pereyra, and A. C. Carciofi}
\affil{Instituto de Astronomia, Geof\'\i sica e Ci\^encias Atmosf\'ericas, Universidade de S\~ao Paulo, Rua do Mat\~ao, 1226, S\~ao Paulo, SP 05508-900, Brazil, e-mail: mario@astro.iag.usp.br}

\begin{abstract}
We discuss the nature of the circumstellar envelopes around the B[e] supergiants (B[e]SG) in the Magellanic Clouds (MC). Contrary to those in the Galaxy, the MC B[e]SG 
have a well defined luminosity and can be considered members of a well defined
class. We discuss spectroscopy and optical broadband polarimetry and spectropolarimetry data. These data show for the first time detailed changes in the polarization across several spectral features.

We show that the envelopes of the B[e]SG are generally variable. Broadband polarimetry data show that the envelopes are definitely non-spherically symmetric and large non-axisymmetric ejections may occur. In addition to that, spectropolarimetry is coming of age as a tool to study the B[e]SG envelope structure.
\end{abstract}


\section{Introduction}
\label{introduction}

A non-spherical wind model, namely, the two-component model of Zickgraf and collaborators \citep{zic85}, has become a paradigm for the circumstellar environment of the B[e] supergiants. The Magellanic B[e]SG are therefore important since, contrary to their Galactic counterparts, their luminous supergiant nature is unquestioned \citep{zic90}.

Polarimetry provides insight into the physical mechanisms operating in and around the source as well as the source's asymmetry . A particularly attractive feature of the technique is that the source does not have to be angularly resolved. This is often the case for stellar envelopes, especially outside the Milky Way, and the study of the MC B[e]SG thus benefits from polarimetric observations.

In the context of circumstellar envelopes mechanisms that polarize light include Rayleigh scattering from atoms and molecules, Mie (i.e., dust) scattering and Thompson scattering by free electrons. A spherical envelope of any such scatterers produces no net polarization; an observed non-zero polarization (after a possible interstellar polarization subtraction) signals the presence of a non-spherical envelope.

An example is a hot star surrounded by an unresolved, ionized, non-spherical envelope, such as those around Be and B[e] stars. The observer receives direct (i.e., mostly unpolarized) stellar flux as well as flux which has been scattered (hence polarized to some extent) within the envelope. The observed degree of polarization is essentially the ratio between the scattered and total flux. \citet{mag92} has shown that the MC B[e]SG do present intrinsic polarization, consistent with the non-spherical envelopes inferred from spectroscopic observations of \citet{zic85}. An important thing to note is that the polarized flux carries information about the emission and absorption processes that occur in the envelope. One such case is the Be star $\zeta$ Tau \citep{bjo92, bjo06} which shows line absorption by Fe II and Fe III barely noticeable in the total flux but quite evident in the polarization spectrum. 

The net continuum polarization resulting from an envelope of Rayleigh or Thomson scatterers has been modeled by \citet{dan80}, \citet{cas87} and \citet{woo96}. Since there is no phase shift involved, there is no circular polarization produced by Thomson or Rayleigh scattering. On the other hand, dust particles will produce circular polarization from multiple scattering. Examples of dust scattering envelopes have been presented by \citet{dan82}.

\section{Techniques}
\label{Instrumentation}

In addition to the first Stokes parameter, the total Flux I, the use of polarimetric optics ahead of cameras and spectrographs \citep[e.g., ][]{mag96, kay99} allows one to measure the fractional values of the three remaining parameters, $q=Q/I$, $u=U/I$ and $v=V/I$. $Q$ provides how much of the flux vibrates at 0$\raisebox{1ex}{\scriptsize o}$ (i.e., towards North) over that at 90$\raisebox{1ex}{\scriptsize o}$,  $U$ does the same for 45$\raisebox{1ex}{\scriptsize o}$ and 135$\raisebox{1ex}{\scriptsize o}$ and $V$ describes how much of the flux is right-circularly polarized over that which is left-circularly polarized. The ${Q, U}$ pair thus describes linear polarization and they relate to the polarization degree and position angle (PA) by P=$\sqrt{(Q^{2}+U^{2})}$ and 
${2\theta}=tan^{-1}(U/Q)$. In a ${Q- U}$ diagram then, a point of cartesian coordinates $(Q, U)$ has polar coordinates ($P$, $2\theta$).

\begin{figure}[!ht]
%\vspace{6cm}
%\plottwo{Plambda.eps}{Ptime_ed.eps}
\plotone{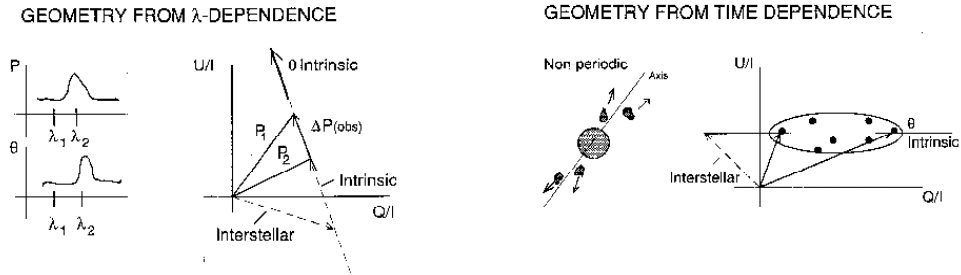}
\caption{Schematic Q-U plots showing polarization values at two distinct wavelengths (left) and at different times (right). Adapted from \citet{nor96}. \label{QUPol}}
\end{figure}

Since {q, u} are often $\ll$1, these parameters are the ones most often plotted in the useful $Q-U$ diagram. In the case of spectropolarimetry, the left diagram in Figure~\ref{QUPol} shows how the continuum-line difference provides the intrinsic polarization plane even without knowledge of the foreground interstellar polarization. The right diagram on Figure~\ref{QUPol} shows the case where the star has stochastic ejections of blobs or material preferentially in and around a plane. If no preferential ejection plane existed in the latter case, then the intrinsic points would be scattered around an interstellar value (if there were no underlying disk, say, in addition to the blob enhancements).

The broadband polarimetric observations reported here have been carried out with the IAG-USP 60cm and LNA 1.60m telescopes, at the LNA observatory in Brazil, and the 1.5m telescope at CTIO, Chile. Instrumental and data reduction details can be found in \citet{pm02, pm04}. The spectropolarimetric observations have been carried out at the CTIO's Blanco 4m telescope. Data reduction was done with an IRAF spectropolarimetric package written by us at IAG. We also report on some of the echelle spectra obtained with the ESO 1.5m telescope and the FEROS spectrograph.

\section{Broadband Polarization of B[e]SG in the Magellanic Clouds}
\label{broadbandpol}

The first observations of the optical polarization of B[e]SG in the MC have been reported by \citet{mag92}. He detected intrinsic polarization in several of them and showed that the data were consistent with the MC B[e]SG having non-spherically symmetric envelopes with a range of intrinsic polarizations. They were also consistent with the spectroscopic data: stars viewed nearly edge-on showed the largest polarizations.

The polarization data seemed then to correlate best with the near infrared (IR) excess than with the average envelope electron density. The latter correlation was somewhat 'spoiled' by AV 16, shown later to be binary \citep{zic96}. Excluding this star, \citet{mel01} showed that the polarization correlated better with the electron density than IR excess. This suggests that Thomson scattering is the main polarizing mechanism in B[e] SG envelopes.

\citet{mel01} modeled the continuum polarization of the Magellanic B[e]SG. They found that the electron density distribution is closer to being homogeneous rather than with an $r^{-2}$ dependence. Interestingly, the data are best fitted by a spherical envelope  in which the density decreases smoothly from the equatorial disk towards the poles \citep{wat87} rather than with a cylindrical distribution, thus supporting the two-component wind model of the B[e] supergiants.

\begin{figure}[!ht]
%\vspace{6cm}
\plottwo{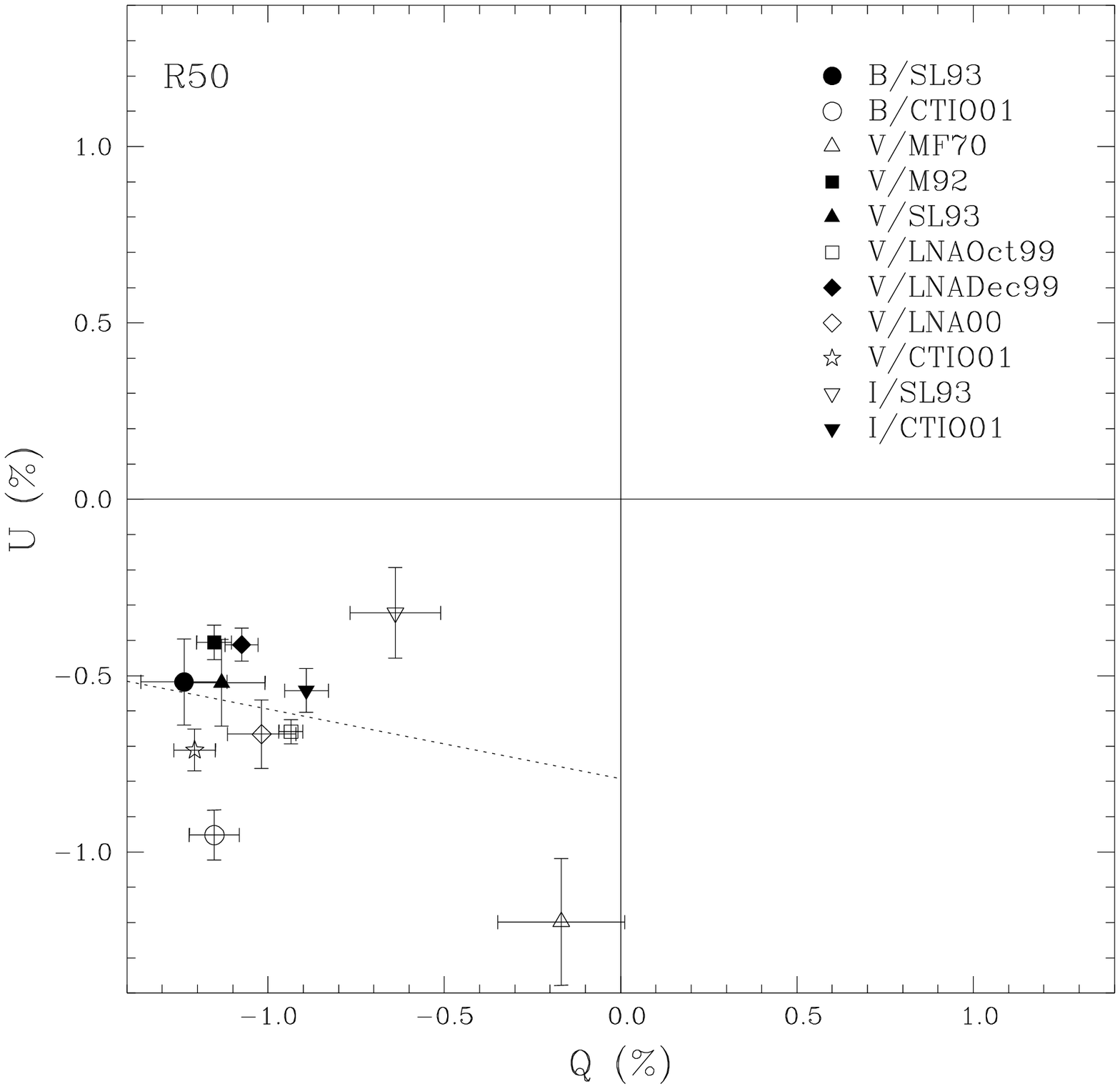}{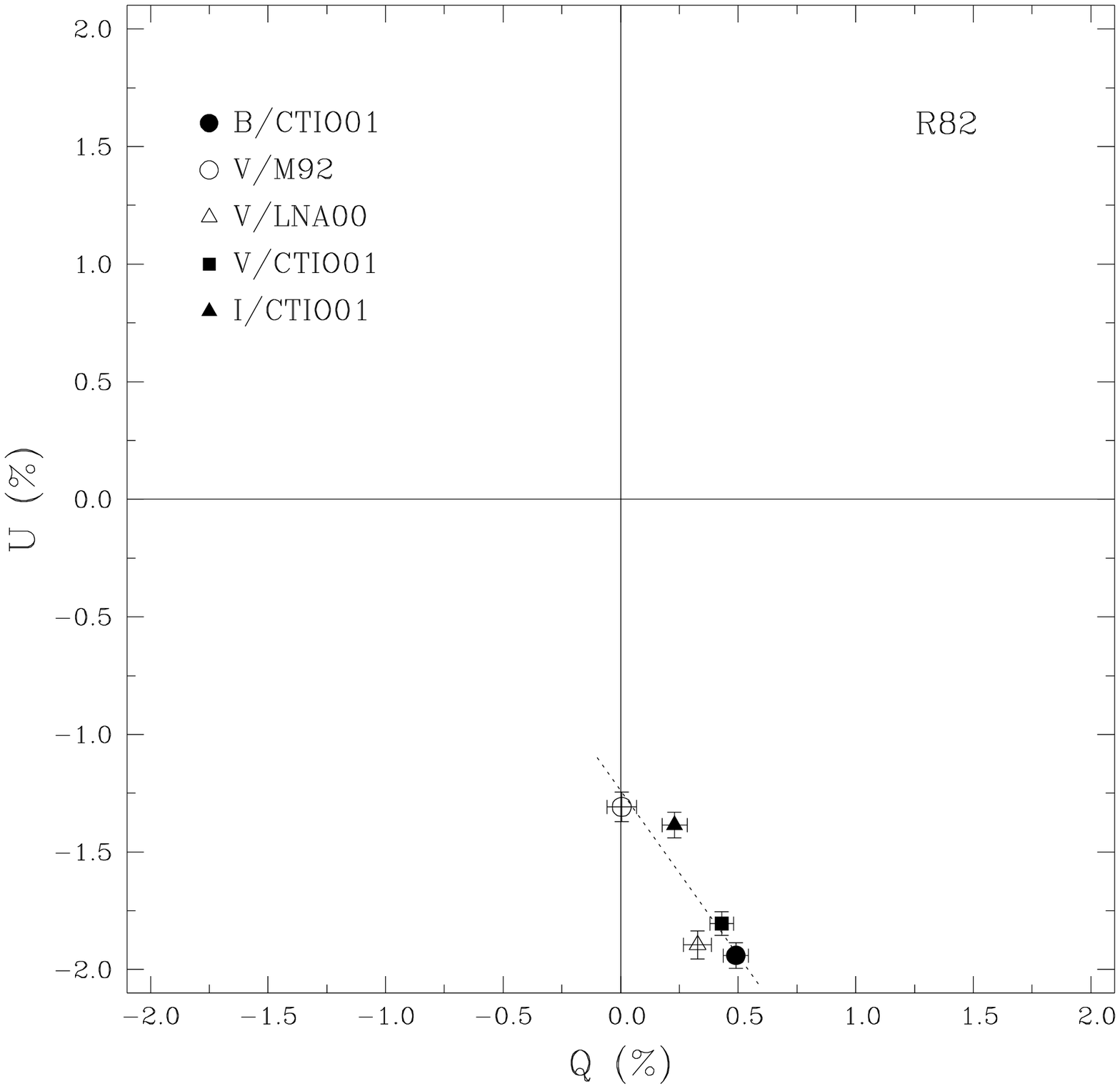}
\caption{Normalized Q-U plots for R50 (left) and R82 (right) \citep{mel04}.\label{ImPol}}
\end{figure}

\citet{mel04} presented broadband data for several B[e]SG obtained from October 1999 through October 2002. Figure \ref{ImPol} shows the Q-U diagram for R50, an SMC B[e]SG. From field stars, \citet{mel04} estimated the interstellar component as $(0.39\pm0.03)$\% at $(104\pm3)\raisebox{1ex}{\scriptsize o}$, consistent with Fig. \ref{ImPol}. We also see that the polarization values do not lie along the dashed line, indicating that, despite seen equatorially, non-axisymmetric ejections do occur in R50's envelope. Using the results of modeling blobs in hot star winds of \citet{rod00}, these data immediately tell us that these ejected blobs are comparable in size to the star and are located near the base of the envelope.

This contrasts with R82, an equatorially seen B[e]SG (Fig. \ref{ImPol}). While the polarization varies, suggesting overall density changes, the envelope geometry has remained stable during the observations, with the data along a line.

\section{Spectroscopy of B[e]SG in the Magellanic Clouds}
\label{spectra}

%\begin{figure}[!ht]
%\vspace{6cm}
%\plottwo{s18halfa_ed.eps}{s18heii_ed.eps}
%\caption{Normalized flux profiles for S18 at $H\alpha$ (left) and HeII 4686A (right) in 2001 and 2002 \citep{mel04}.\label{S18}}
%\end{figure}

Echelle spectra of half a dozen MC B[e]SG were taken in November 2001 and four consecutive nights in September 2002 \citep{mel04}. R 50 and S22 showed variable $H\alpha$ emission from 2001 to 2002, while R82, our equatorial, well-behaved friend (sec. \ref{broadbandpol}) showed no sign of variability in the period.

In the pole-on SMC object S18, in 2001 $H\alpha$ and $H\beta$ showed P-Cygni profiles while in 2002 a more pure emission was present, indicating a larger ionized volume in the envelope \citep{mel04}. This is confirmed by the HeII 4686A line, which was absent in 2001 and strong and variable in 2002. S18 might be a binary star \citep{sho87, zic89} and our data are consistent with a hotter companion interacting with the B[e]SG envelope at times.

\section{Spectropolarimetry of B[e]SG in the Magellanic Clouds}
\label{specpol}

\begin{figure}[!ht]
%\vspace{6cm}
\plottwo{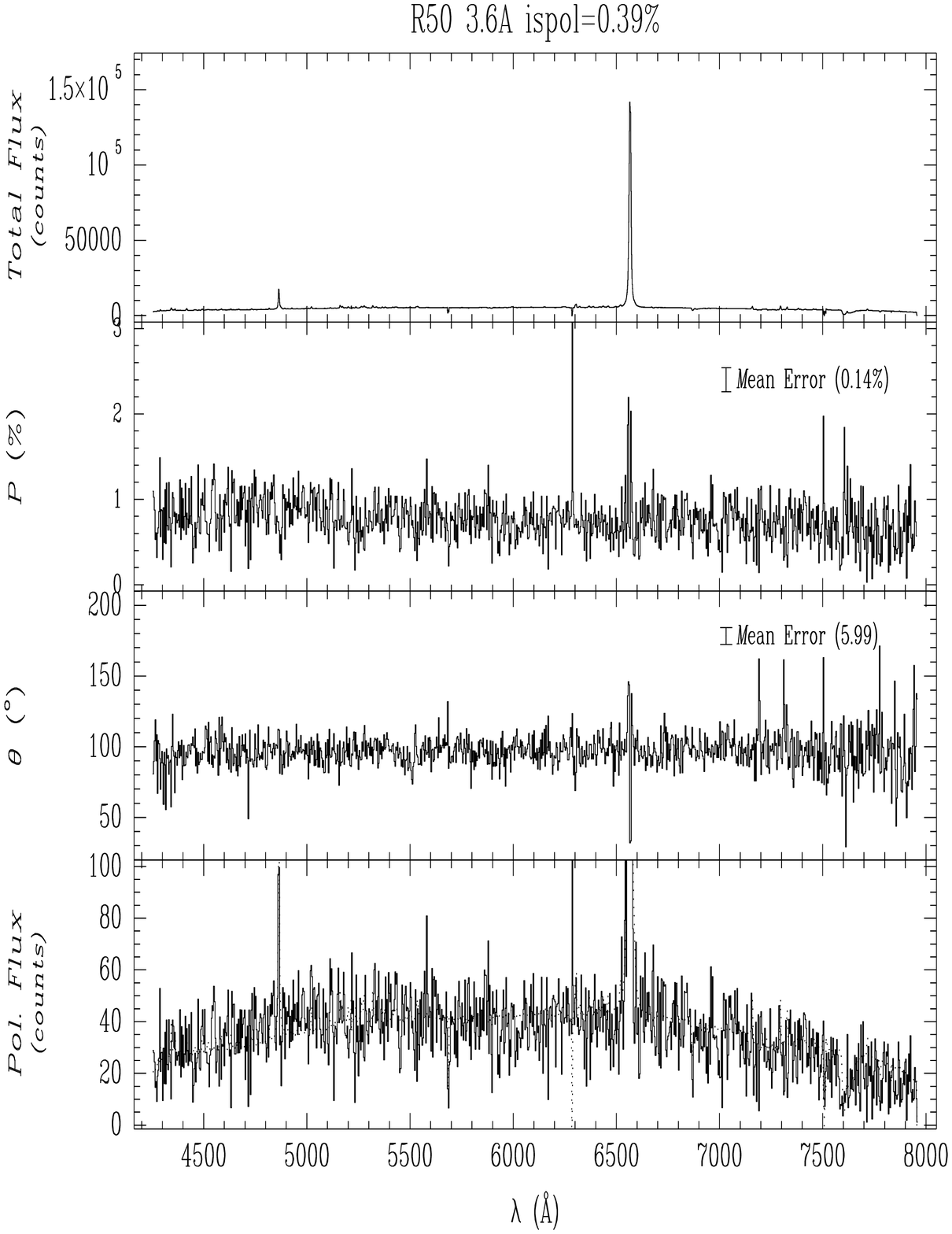}{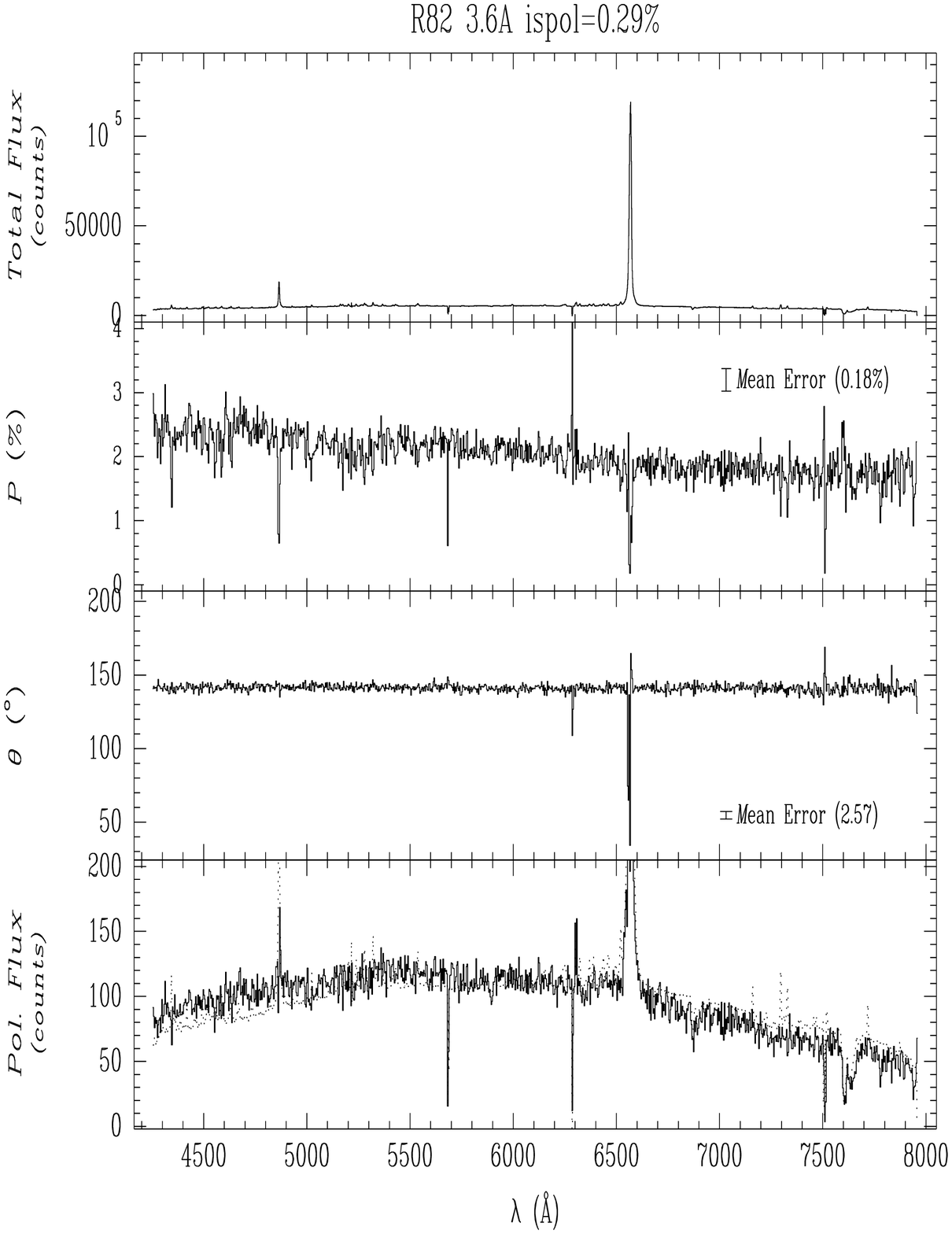}
\caption{Spectropolarimetry data for R50 (left) and R82 (right) \citep{mag06}. The original data were corrected for interstellar polarization and binned by 3 pixels (3.6A). In each plot, from top to bottom panels: total flux, percent polarization, polarization position angle and polarized flux. The (scaled down) total flux is overplotted on the polarized flux panel as a dotted line. Features near 6300A and 7500A are detector artifacts.\label{ContPol}}
\end{figure}

Spectropolarimetry should be an important tool to probe the MC B[e]SG envelopes \citep{mag92} since the scattering opacity (which produces polarization) competes with line emission and absorption opacities (which tend to suppress polarization) as a function of wavelength and throughout the envelope. An example is when the polarization is mostly produced near the star and the thermal, unpolarized line emission flux comes from a more extended region, such as is often the case for $H\alpha$. The line photons have a lower scattering opacity before leaving the envelope and the observed polarization across the line decreases. S22 \citep{rsl93}, the only MC B[e]SG with published spectropolarimetry thus far, shows this very picture. We have recently obtained spectropolarimetric observations of R50, R82, R126, S12 and S22 \citep{mag06} and we present here a sample of the preliminary results.

\begin{figure}[!ht]
%\vspace{6cm}
\plottwo{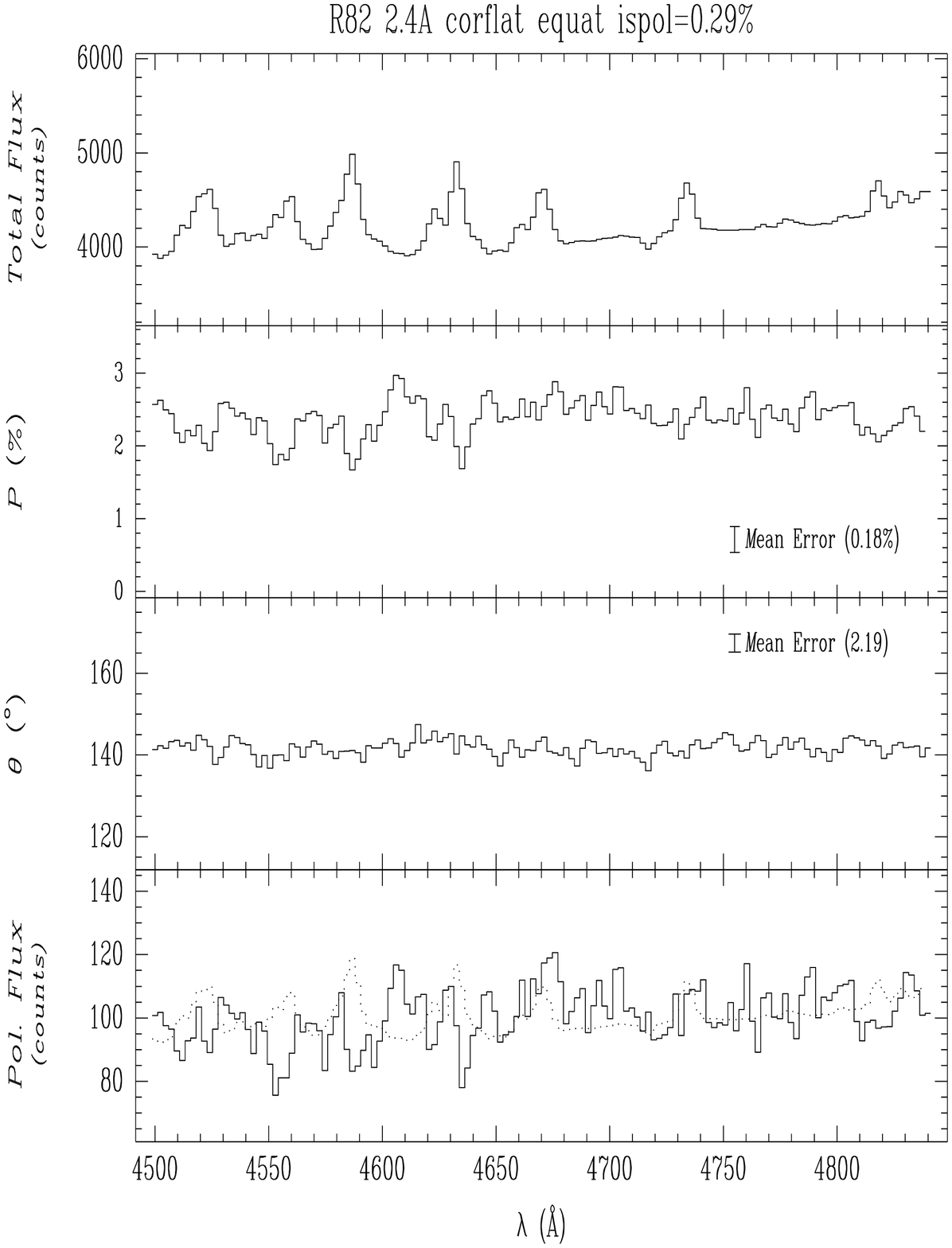}{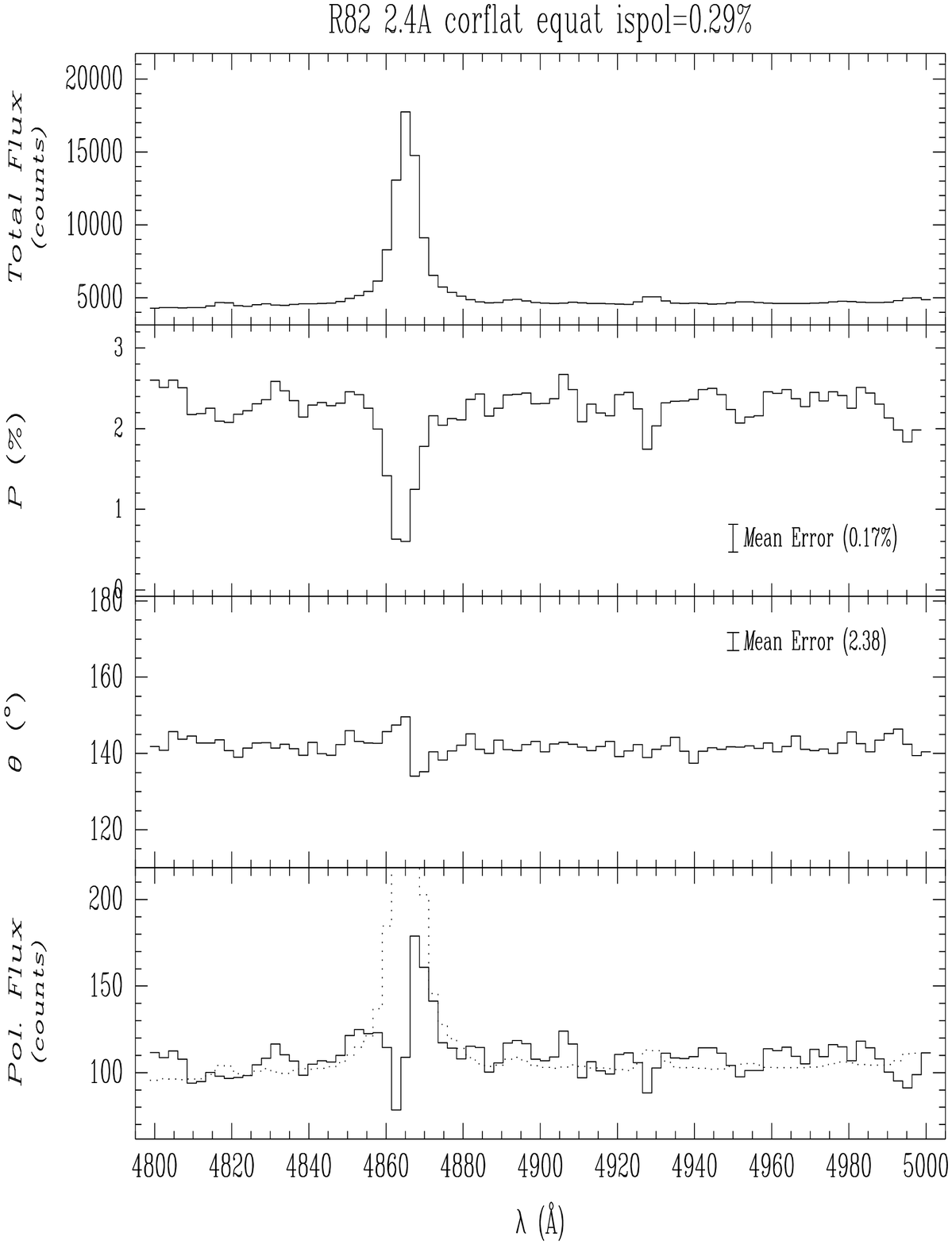}
\caption{Spectropolarimetric data for R82 around (right) and shortwards (left)  of $H\beta$ \citep{mag06}. The data were corrected for interstellar polarization and binned by 2 pixels (2.4A). Panels are as in Fig. \ref{ContPol}. \label{LinePol}}
\end{figure}

\subsection{Continuum data}
\label{Continuum}

Figure \ref{ContPol} shows our polarization data for R50 and R82 with the interstellar polarization removed, reflected by the zero-slope PA. Their polarization shows a 'non-white' dependence, which would otherwise result if only electron scattering was operative. This can be seen more clearly in the lower panel for R82, where the polarized flux (Pflux = polarization $\times$ flux) is bluer than the direct flux. The Pflux can be roughly understood as the spectrum of the source as seen by the scatterers.

These observations suggest that either dust and/or hydrogen continuum opacity (cf. sec. \ref{introduction}) may be playing  a role. This will have to be quantified by detailed modeling, in progress. Observationally, the detection of circular polarization in B[e]SG would support scattering by dust. Near IR polarimetry would be also helpful in evaluating the role of dust scattering.

\subsection{Line effects}
\label{Line effects}

Figure \ref{LinePol} shows two expanded spectral sections for R82. Around $H\beta$, the polarization changes dramatically across the line. The decrease across the line is not a simple addition of unpolarized line photons out in the envelope. The polarized flux, in the bottom panel, has a P-Cygni profile indicating the velocity structure of the inner part of envelope, which is being scattered further out. There is also an indication of a PA change across the line: when we resolve the line, the photons come from different parts of a rotating, expanding envelope \citep{woo93, hil94, hil96, vin05}, part of which may be occulted by the star. 

Also on the right panel, we see the much weaker (compared to $H\beta$) [FeII] 4927A line which shows a decrease in the polarization across the line with little PA change. This would be expected in a line that is formed outside where most of the polarization is originating. The line is much weaker in the Pflux spectrum, as expected from dilution by unpolarized flux but a more detailed analysis is needed. The FeII 4586A \& 4632A lines on the left panel show a similar behavior.

\section{Conclusions and Perspectives}

Broadband optical polarization data indicate that the MC B[e]SG have non-spherical envelopes where both axisymmetrical and non-axisymmetric ejections may occur. Electron scattering seems to be the main polarizing mechanism. However dust and/or H opacity may play a role as some objects show a non-white polarization spectrum. If detected, circular polarization would favor dust.

Broadband polarization monitoring should provide more details on the stability of the envelopes and the non axi-symmetric ejections. Plans exist for a 2m class Robotic Telescope in the Chilean Andes, with imaging and optical/IR spectroscopy capabilities. Interested partners are encouraged to contact us.

Spectropolarimetry, including the data reported here, shows great power for studying the line formation regions and the structure of the B[e]SG, providing feedback to and building upon models such as those by \citet{bj06}, \citet{car06}, \citet{kra06} and \citet{kra05}. Such modeling, together with high signal-to-noise spectropolarimetry of the MC B[e]SG with the VLT and the forthcoming SALT telescopes, will allow spectropolarimetry to realize its full potential.

\acknowledgments{We gratefully acknowledge support by FAPESP grant 01/12589-1. AMM is indebted to partial support by FAPESP and CNPq. RM acknowledges support by CNPq. AP and AC are supported by FAPESP. We thank Laura Kay and Hien Tran for support during the CTIO spectroscopic run as well as Karen Bjorkman and John Wisniewski for support during the CTIO imaging runs. Regina Schulte-Ladbeck is gratefully acknowledged for providing the 1993 data points of R50 in Figure \ref{ImPol}. We thank Ken Nordsieck for the original diagram of Fig. \ref{QUPol}.}

\end{document}